\documentclass[twocolumn,preprintnumbers,amsmath,amssymb,pra,a4paper]{revtex4}

\usepackage{graphicx}
\usepackage{dcolumn}
\usepackage{bm}
\usepackage{braket}
\usepackage{mathbbol}

\begin{document}

\title{Capacity of a bosonic memory channel with Gauss-Markov noise}

\author{Joachim Sch\"afer} 
\author{David Daems}
\author{Evgueni Karpov} 
\affiliation{QuIC, Ecole Polytechnique, CP 165, Universit\'e Libre de
Bruxelles, 1050 Brussels, Belgium}

\author{Nicolas J. Cerf} 
\affiliation{QuIC, Ecole Polytechnique, CP 165, Universit\'e Libre de
Bruxelles, 1050 Brussels, Belgium}
\affiliation{Research Laboratory of Electronics, Massachusetts Institute of Technology, Cambridge, MA 02139}

\begin{abstract}
We address the classical capacity of a quantum bosonic memory channel with additive noise, subject to an input energy constraint. The memory is modeled by correlated noise emerging from a Gauss-Markov process. Under reasonable assumptions, we show that the optimal modulation results from a ``quantum water-filling'' solution above a certain input energy threshold, similar to the optimal modulation for parallel classical Gaussian channels. We also derive analytically the optimal multimode input state above this threshold, which enables us to compute the capacity of this memory channel in the limit of an infinite number of modes. The method can also be applied to a more general noise environment which is constructed by a stationary Gauss process. The extension of our results to the case of broadband bosonic channels with colored Gaussian noise should also be straightforward.
\end{abstract}

\maketitle

\section{Introduction}   

The growing importance of quantum communication motivates a strong research activity on quantum channels. The classical capacity of a communication channel is defined as the supremum of the rate of classical information that can be transmitted via the channel. An important question for quantum channels is whether or not entangling the channel inputs may improve the transmission rate. The additivity conjecture proven to be true for some memoryless quantum channels (see \cite{hiro06} and references herein) implies that entanglement does not improve the transmission rate. Although this conjecture was disproved recently for another class of channels in Ref.~\cite{hast09}, it remains an interesting question in itself to determine whether entanglement helps or not to achieve the capacity. In particular, it was shown for some channels with \textit{memory}, where the environment exhibits a correlation be- tween subsequent uses of the channel, that the transmission rate can be increased by using entangled input states.

In the case of a quantum memory channel with discrete alphabet, the first studies of the classical capacity considered a depolarizing channel \cite{macc02} and a quasiclassical depolarizing channel \cite{macc04} and showed the existence of a threshold on the degree of memory above which entangled signals improve the transmission rate with respect to product states. Further studies \cite{bowe04,bowe05} on a qubit channel with finite memory derived bounds on the classical and quantum capacities. In Ref.~\cite{kret05}, the classical and quantum capacities were discussed in a general framework, where furthermore, a malicious third party may have control of the initializing memory. The case of general Pauli channels with memory was studied explicitly in Ref.~\cite{daem07}, where it was shown that the optimal states are either product states or Bell states separated by a memory threshold. The behavior of channels with correlated error was connected to quantum phase transitions in many-body systems in Ref.~\cite{plen07}. The quantum capacity of a dephasing channel was computed in Ref.~\cite{arri07}, where an enhancement of the capacity with increasing memory was proven. For higher dimensions, it has been shown that the capacity of qudit channels exhibits the same threshold phenomenon as Pauli qubit channels \cite{karp06,kari06}.

For a quantum channel with continuous alphabet, it was first shown for an additive bosonic channel and a lossy bosonic channel, respectively, \cite{cerf05,giov05} that in the presence of a memory, some degree of entanglement between the input states is necessary to achieve the capacity, which is in contrast to the behavior reported for discrete quantum channels. Indeed, the optimal input states correspond to Einstein- Podolski-Rosen (EPR) states with finite squeezing which in- creases with the degree of memory, in contrary to either maximally or nonentangled states in the discrete case. A lossy bosonic channel with correlated environment and an input energy constraint has been studied \cite{rugg05,pily08,lupo09}. Lower and upper bounds for the classical capacity were derived in Ref.~\cite{giov05} and the capacity was calculated in Refs.~\cite{rugg05,pily08,lupo09}.

In this paper, we discuss a bosonic memory channel with additive noise modeled by a Gauss-Markov process. We present an extension of the model of Ref.~\cite{cerf05}, where two uses of a bosonic additive channel with correlated noise were treated, to the case of $n$ channel uses with Gauss-Markov noise. We determine the classical capacity of the channel in the asymptotic limit.

In Sec. \ref{sec:boschan}, we recall the definition of the classical capacity of quantum channels and specify the model of the channel under an input energy constraint. Section III is devoted to the treatment of a monomodal channel with phase-dependent noise where the quantum water-filling emerges as optimal solution. In Sec. IV, we introduce the method for finding the capacity and apply it to the Gauss-Markov memory. Under reasonable assumptions, we determine the optimal solution as a global water-filling solution and obtain the classical capacity of the channel. Section V treats the capacity in several limiting cases: the limit of full correlations, the classical limit, the case of a modified thermal noise, and the transition from finite to infinite uses.

\section{\label{sec:boschan}Bosonic additive channel}

In order to send classical information through a quantum channel one chooses an alphabet and associates its letters to quantum states $\rho^\mathrm{in}_i$, where 
index $i$ labels the letters of the alphabet. The input states sent through the channel interact with the environment and thus are modified at the output. The action of the channel $T$ is a completely positive, trace-preserving map acting on the input states $\rho^\mathrm{in}_i$: 
\begin{equation}\label{eq:channel}
  \rho^\mathrm{out}_i=T[\rho^\mathrm{in}_i].
\end{equation} 
In the messages, each letter appears with a certain probability $p_i$ so that the overall modulated input state is described as a mixture
\begin{equation}\label{eq:mix}
  \rho^\mathrm{in} = \sum_i p_i \rho^\mathrm{in}_i.
\end{equation} 
By linearity of $T$, Eq.~(\ref{eq:channel}) determines as well the action of the channel on the overall modulated input (\ref{eq:mix}), that is,
\begin{equation}
	\overline{\rho} \equiv \sum\limits_i{p_i \rho^\mathrm{out}_i} = T[\rho^\mathrm{in}],
\end{equation}
where we will refer to $\overline{\rho}$ as the overall modulated output. In order for the overall modulated input to be physical it has to obey the energy constraint
\begin{equation}
  \sum\limits_i{p_i \, \mathrm{Tr}(\rho_i^\mathrm{in} \, a^\dagger a)} \leq \overline{n},
  \label{eq:enconstr}
\end{equation}
where $\overline{n}$ is the maximum mean photon number per use of the channel and will be referred to as ``input energy'' in the following. 

The classical capacity $C(T)$ of the channel $T$ represents the supremum on the amount of classical bits which can be transmitted per invocation of the channel via quantum states
in the limit of an infinite number of channel uses. This quantity can be calculated with the help of the so-called \textit{one-shot} capacity, defined as in \cite{schu97}:
\begin{equation}
  C_1(T) = \sup_{\rho_i^\mathrm{in},p_i}{\; \chi},
  \label{eq:oscapacity}
\end{equation}
where the Holevo $\chi$-quantity reads
\begin{equation}
  \chi = S\left(\sum_i{p_i \, T[\rho_i^\mathrm{in}]}\right) - \sum\limits_i{p_i \, S(T[\rho_i^\mathrm{in}])},
  \label{eq:holevo}
\end{equation}
with the von Neumann entropy $S(\rho) = -\mathrm{Tr}(\rho\log{\rho})$ where $\log$ denotes the logarithm to base 2. The supremum in \eqref{eq:oscapacity} is taken over all ensembles of $\{p_i,\rho_i^\mathrm{in}\}$ of probability distributions $p_i$ and pure input ``letter'' states  $\rho_i^\mathrm{in}$ \cite{schu97}.

The term ``one-shot'' means that only one invocation of $T$ is needed to calculate Eq.~(\ref{eq:oscapacity}). Using this quantity the capacity $C(T)$ defined as above may be evaluated in the following way. A number $n$ of consecutive uses of the channel $T$ can be equivalently considered as one parallel $n$-mode channel $T^{(n)}$, which is used only one time. Then the capacity $C(T)$ is evaluated in the limit:
\begin{equation}
  C(T) = \lim_{n \rightarrow \infty} \frac{1}{n} C_1(T^{(n)}).
\end{equation}

Let us now assume $T^{(n)}$ to be a $n$-mode bosonic additive channel with memory. In the following, the number of modes of this channel corresponds to the number of mono-modal channel uses. Each mode $j$ is associated with the annihilation and creation operators $a_j,a_j^{\dagger}$, respectively, or equivalently to the quadrature operators $q_j = (a_j + a_j^{\dagger})/\sqrt{2},p_j = i(a_j - a_j^{\dagger})/\sqrt{2}$ which obey the canonical commutation relation $[q_i,p_j] = i\delta_{ij}$, where $\delta_{ij}$ denotes the Kronecker-delta. By ordering the quadratures in a column vector 
\begin{equation}
	\bm{R} = (q_1,...,q_n;p_1,...,p_n)^\mathrm{T},
\end{equation}
we can define the displacement vector $\bm{m}$ and covariance matrix $\bm{\gamma}$ of an $n$-mode state $\bm{\rho}$ as
\begin{equation}
	\begin{split}
	\bm{m} & = \mathrm{Tr}{[ \, \bm{\rho}  \bm{R}]}\\
	\bm{\gamma} & = \mathrm{Tr}{[ (\bm{R} - \bm{m}) \, \bm{\rho} \, (\bm{R} - \bm{m})^{\ensuremath{\mathsf{T}}} ]} - \frac{1}{2}\bm{J},
	\end{split}
\end{equation}
where 
\begin{equation}
	\bm{J} = i\begin{pmatrix} 0 & \mathbb{1}\\-\mathbb{1} & 0 \end{pmatrix}
	\label{eq:J}
\end{equation} 
is the symplectic or commutation matrix with the $n \times n$ identity matrix $\mathbb{1}$. In this paper we focus on Gaussian states, which are fully characterized by $\bm{m}$ and $\bm{\gamma}$. Furthermore, without loss of generality, we set the displacement of overall modulated states and the means of classical Gaussian distributions to zero, because displacements do not change the entropy.

For the bosonic channel, the encoding of classical information is made according to a continuous alphabet, where the previous discrete letter with index $i$ is replaced by the real and imaginary part of a complex number $\alpha$. A message of length $n$ is therefore encoded in a $2n$ real column vector\\ $\bm{\alpha} = (\Re{\{\alpha_1\}},\Re{\{\alpha_2\}},...,\Re{\{\alpha_n\}},\Im{\{\alpha_1\}},...,\Im{\{\alpha_n\}})^{\ensuremath{\mathsf{T}}}$. Physically, this encoding corresponds to a displacement of the $n$-partite Gaussian input state in the phase space by $\bm{\alpha}$ and is denoted by $\bm{\rho^\mathrm{in}_\alpha}$. The Wigner function of $\bm{\rho^\mathrm{in}_\alpha}$ reads
\begin{equation}
  W^\mathrm{in}_{\bm{\alpha}}(\bm{R}) = \frac{\exp{[-\frac{1}{2}(\bm{R}-\sqrt{2}\bm{\alpha})^\dagger \, \bm{\gamma_\mathrm{in}}^{-1} \, (\bm{R}-\sqrt{2}\bm{\alpha})]}}  {(2\pi)^{n} \sqrt{\det{(\bm{\gamma_\mathrm{in}})}}}. 
\end{equation} 
Throughout this work, we use the recently proven conjecture that a coherent state minimizes the entropy of a mono-modal Gaussian thermal channel \cite{lloy09}. As we show in Secs. \ref{sec:mono} and \ref{sec:mark}, we only need the extension of this proof to the case of a mono-modal Gaussian channel with anisotropic noise to justify that the optimal $n$-partite input state is Gaussian. As a result we only consider Gaussian distributions of the letters in the messages so that the overall modulated input state sent through the channel is a Gaussian mixture
${\bm\rho^{\mathrm{in}}}=\int d^{2n}\bm{\alpha}f({\bm{\alpha}}) {\bm{\rho^{\mathrm{in}}_\alpha}}$, where $d^{2n}\bm{\alpha} = d\Re{\{\alpha_1\}}d\Im{\{\alpha_1\}}...d\Re{\{\alpha_n\}}d\Im{\{\alpha_n\}}$ with Gaussian distribution
\begin{equation}
  f({\bm \alpha}) = \frac{\exp{[-\bm{\alpha}^{\ensuremath{\mathsf{T}}} \bm{\gamma_\mathrm{mod}}^{-1} \bm{\alpha}]}}{\pi^n \sqrt{\det{(\bm{\gamma_\mathrm{mod}})}}},
  \label{eq:mod}
\end{equation} 
centered at zero and characterized by the modulation covariance matrix $\bm{\gamma_\mathrm{mod}}$.

As we are no longer dealing with probability distributions $p_i$ but with probability densities $f(\bm{\alpha})$, the summations in the formulae above are replaced by proper integrations. The action of $T^{(n)}$ on an input state carrying a message $\bm{\alpha}$ reads as in \cite{cerf05}
\begin{equation}
  \begin{split}
         & T^{(n)}[\bm{\rho^{\mathrm{in}}_\alpha}] = \bm{\rho^\mathrm{out}_\alpha} = \int d^{2n} {\bm\beta} \, f_\mathrm{env}({\bm\beta})\\
 					 & \times  D(\beta_n) \otimes ... \otimes D(\beta_1) \; \bm{\rho^\mathrm{in}_\alpha} \; D^{\dagger}(\beta_1) \otimes ... \otimes D^{\dagger}(\beta_n),
  \end{split}
  \label{eq:changen}
\end{equation} 
with $d^{2n}\bm{\beta} = d\Re{\{\beta_1\}}d\Im{\{\beta_1\}}...d\Re{\{\beta_n\}}d\Im{\{\beta_n\}}$, $\bm{\beta} = (\Re{\{\beta_1\}},...,\Re{\{\beta_n\}},\Im{\{\beta_1\}},...,\Im{\{\beta_n}\})^{\ensuremath{\mathsf{T}}}$ and the displacement operator $D(\beta_j) = e^{\beta_j \hat a_j^\dagger - \beta_j^{*} \hat a_j}$. The displacement is applied according to the Gaussian distribution of the environment 
\begin{equation}
	f_\mathrm{env}({\bm \beta}) = \frac{\exp{[-\bm{\beta^{\ensuremath{\mathsf{T}}}} \, \bm{\gamma_\mathrm{env}}^{-1} \, \bm{\beta}]}}{\pi^n \sqrt{\det{(\bm{\gamma_\mathrm{env}})}}}, 
\end{equation}
with (classical) covariance matrix $\bm{\gamma_\mathrm{env}}$. If this matrix is not diagonal, then the environment introduces correlations between the successive uses of the channel. These correlations model the memory of the channel.

Since we centered the distributions of the environment and modulation as well as the Wigner function of the input state around zero, the covariance matrices of the state carrying the message $\bm{\alpha}$ at the output of the channel $\bm{\rho^\mathrm{out}_\alpha}$ and of the overall modulated output state $\bm{\overline{\rho}}$, read, respectively,
\begin{equation}
  \begin{split}
    \bm{\gamma_\mathrm{out}} = \bm{\gamma_\mathrm{in}} + \bm{\gamma_\mathrm{env}}\\
    \bm{\overline{\gamma}} = \bm{\gamma_\mathrm{out}} + \bm{\gamma_\mathrm{mod}}.
  \end{split}
  \label{eq:chancov}
\end{equation}
The one-shot capacity of such a system is
\begin{equation}
  C_1(T^{(n)}) = \sup_{\bm{\gamma_\mathrm{in}},\bm{\gamma_\mathrm{mod}}}{\chi_n},
  \label{eq:ncapacity}
\end{equation}
with the Holevo $\chi$-quantity which then reduces to
\begin{equation}
  \chi_n = S(\bm{\overline{\rho}}) - S(\bm{\rho^\mathrm{out}_\alpha}).
  \label{eq:nchi}
\end{equation}

In the case of a Gaussian state $\bm{\rho}$, the von Neumann entropy can be expressed in terms of the symplectic eigenvalues $\nu_j$ of its covariance matrix:
\begin{eqnarray}
	S(\bm{\rho}) & = & \sum_j{g\left(|\nu_j|-\frac{1}{2}\right)} \\
	g(x) & = & \left\{ \begin{array}{ll} (x+1)\log{(x+1)} - x\log{x} &, \,  x > 0\\
											0 &, \, x = 0.
    				   \end{array}\right.\nonumber
	\label{eq:entropy}
\end{eqnarray}
The energy (or mean photon number) constraint \eqref{eq:enconstr} is given by
\begin{equation}
   \frac{1}{2n}\left( \mathrm{Tr}\left({\bm{\gamma_\mathrm{in}}}\right) + \mathrm{Tr}\left({\bm{\gamma_\mathrm{mod}}}\right)\right) - \frac{1}{2} = \overline{n}.
   \label{eq:enconstrcov}
\end{equation}

We remark that in Eqs. (\ref{eq:nchi}-\ref{eq:entropy}) the quantities depend only on the covariance matrices and not on a particular message $\bm{\alpha}$. Therefore, we can fully discuss the action of the channel solely in terms of covariance matrices.

\section{\label{sec:mono}Mono-modal phase dependent channel}
First we study the mono-modal channel $T$, where the input state undergoes a phase-dependent noise with covariance matrix
\begin{equation}
  \bm{\gamma_\mathrm{env}} =
  \begin{pmatrix}
  \gamma_{\mathrm{env}}^{(q)} & 0\\
	0 & \gamma_{\mathrm{env}}^{(p)}
	\end{pmatrix}, \quad
	\gamma_{\mathrm{env}}^{(q)} \neq \gamma_{\mathrm{env}}^{(p)}.
	\label{eq:monoenv}
\end{equation}
The optimal solution of a channel with such a noise has not yet been fully studied. In Ref. \cite{yen05} the capacity of a multi access channel with such a noise was studied but the optimal input and modulation were not discussed in detail. It was shown \cite{lloy09} for a mono-modal thermal channel that a coherent input state which is modulated according to a Gaussian modulation achieves the capacity. We conjecture that this proof can be extended to the case of the noise given by Eq.~\eqref{eq:monoenv}, i.e. that a Gaussian input state and a Gaussian modulation remain optimal.

The general input and modulation covariance matrices read
\begin{equation}
  \bm{\gamma_\mathrm{in}} = 
  \begin{pmatrix}
  \gamma_\mathrm{in}^{(q)} & \gamma_\mathrm{in}^{(qp)}\\
	\gamma_\mathrm{in}^{(qp)} & \gamma_\mathrm{in}^{(p)}
	\end{pmatrix}, \quad
  \bm{\gamma_\mathrm{mod}} = 
	\begin{pmatrix}
  \gamma_\mathrm{mod}^{(q)} & \gamma_\mathrm{mod}^{(qp)}\\
	\gamma_\mathrm{mod}^{(qp)} & \gamma_\mathrm{mod}^{(p)}
	\end{pmatrix}.
\end{equation}

\begin{figure}
	\centering
		\includegraphics[width=0.3\textwidth]{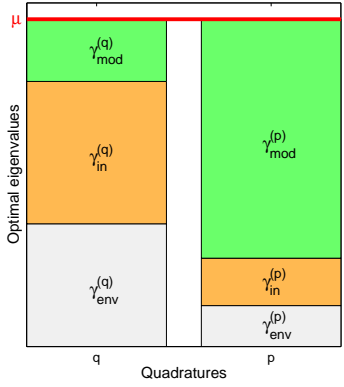} 
	\caption{(Color online) Quantum water-filling solution for the mono-modal channel: Optimal eigenvalues for both quadratures.}
	\label{fig:monqwf}
\end{figure}

We now determine the optimal $\bm{\gamma_\mathrm{in}},\bm{\gamma_\mathrm{mod}}$ by solving equation Eq.~(\ref{eq:ncapacity})
with energy constraint \eqref{eq:enconstrcov} and the requirement to have a pure input, i.e.
\begin{equation}
  \det{\bm{\gamma_\mathrm{in}}} = \frac{1}{4}. 
  \label{eq:monpure}
\end{equation}
With these constraints we can write out the total Lagrangian of the system, i.e.
\begin{widetext}
\begin{equation}
  \begin{split}
  {\cal{L}} = & \, g\left( \sqrt{ \left( \gamma_\mathrm{in}^{(q)} + \gamma_\mathrm{env}^{(q)} + \gamma_\mathrm{mod}^{(q)} \right) \left( \gamma_\mathrm{in}^{(p)} + \gamma_\mathrm{env}^{(p)} + \gamma_\mathrm{mod}^{(p)} \right) - (\gamma_\mathrm{in}^{(qp)} + \gamma_\mathrm{mod}^{(qp)})^2 } - \frac{1}{2} \right)\\ 
  			    - & \, g\left( \sqrt{ \left( \gamma_\mathrm{in}^{(q)} + \gamma_\mathrm{env}^{(q)} \right) \left( \gamma_\mathrm{in}^{(p)} + \gamma_\mathrm{env}^{(p)}\right) - (\gamma_\mathrm{in}^{(qp)})^2 } - \frac{1}{2} \right)\\
  					-  & \, \zeta \left( \gamma_\mathrm{in}^{(q)} + \gamma_\mathrm{mod}^{(q)} + \gamma_\mathrm{in}^{(p)} +  \gamma_\mathrm{mod}^{(p)} \right) - \tau \left( \gamma_\mathrm{in}^{(q)} \gamma_\mathrm{in}^{(p)} \right),
  \end{split}
  \label{eq:monlagrangian}
\end{equation}
\end{widetext}
where $\zeta$ and $\tau$ are Lagrangian multipliers. We arrive at a system of eight equations where the solution reads as follows. First, we find that 
\begin{equation}
  \gamma_\mathrm{in}^{(qp)} = \gamma_\mathrm{mod}^{(qp)} = 0,
\end{equation}
i.e. all three covariance matrices are diagonal in the same basis. Secondly, we obtain
\begin{equation}
  \gamma_\mathrm{in}^{(q)} + \gamma_\mathrm{env}^{(q)} + \gamma_\mathrm{mod}^{(q)} = \gamma_\mathrm{in}^{(p)}+ \gamma_\mathrm{env}^{(p)} + \gamma_\mathrm{mod}^{(p)},
  \label{eq:monthermal}
\end{equation}
which can be regarded as a ``quantum water-filling''\footnote{The ``quantum water-filling'' solution first appeared in the discussion of a different model, i.e. the capacity of the memoryless Gaussian channel in \cite{hole99}, where a classical input signal displaces a quantum noise, which is given by a multimode cavity state.} solution and implies that the optimal output state $\bm{\overline{\gamma}}$ is a thermal state (see Fig.~\ref{fig:monqwf}). This repartition of input energy for the preparation of the quantum state and the classical modulation generalizes the situation of parallel classical channels with a joint energy constraint. Here, we have two channels, one for the $q$ quadrature and one for the $p$ quadrature. 
The eigenvalues of the covariance matrix of the modulated output state read
\begin{equation}
  \overline{\gamma}^{(q)} = \overline{\gamma}^{(p)} = \overline{n} + \frac{\gamma_\mathrm{env}^{(q)} + \gamma_\mathrm{env}^{(p)}}{2} + \frac{1}{2},
  \label{eq:monmixth}
\end{equation}
which simply reflects the total energy of the system. Thirdly, we obtain the optimal degree of squeezing of the input state:
\begin{equation}
  \frac{\gamma_\mathrm{in}^{(q)}}{\gamma_\mathrm{in}^{(p)}} = \frac{\gamma_\mathrm{env}^{(q)}}{\gamma_\mathrm{env}^{(p)}}.
  \label{eq:ratinenv}
\end{equation}
Thus, the squeezing at the input has to match exactly the phase dependence of the noise.

Since all matrices are diagonal we conclude that the variables $\gamma_\mathrm{in}^{(q)},\gamma_\mathrm{in}^{(p)},\gamma_\mathrm{mod}^{(q)},\gamma_\mathrm{mod}^{(p)}$ are the eigenvalues of $\bm{\gamma_\mathrm{in}},\bm{\gamma_\mathrm{mod}}$. Using the condition that the input has to be pure \eqref{eq:monpure} we conclude from Eq.~(\ref{eq:ratinenv}) that
\begin{equation}
  \gamma_\mathrm{in}^{(q)} = \frac{1}{2}\sqrt{\frac{ \gamma_\mathrm{env}^{(q)} }{ \gamma_\mathrm{env}^{(p)} }}, \quad \gamma_\mathrm{in}^{(p)} = \frac{1}{2}\sqrt{\frac{ \gamma_\mathrm{env}^{(p)} }{ \gamma_\mathrm{env}^{(q)} }}.
  \label{eq:monatin}
\end{equation}
Then, the modulation eigenvalues simply read
\begin{equation}
  \begin{split}
    \gamma_\mathrm{mod}^{(q)} & = \mu - \gamma_\mathrm{out}^{(q)}\\  
    \gamma_\mathrm{mod}^{(p)} & = \mu - \gamma_\mathrm{out}^{(p)},
  \end{split}
  \label{eq:monatmod}
\end{equation}
where $\gamma_\mathrm{out}^{(q,p)} = \gamma_\mathrm{in}^{(q,p)} + \gamma_\mathrm{env}^{(q,p)}$ and where the quantum water-filling level $\mu \equiv \overline{\gamma}^{(p)} = \overline{\gamma}^{(q)}$ is determined by Eq.~(\ref{eq:monmixth}). Equations \eqref{eq:monatin} and \eqref{eq:monatmod} are depicted in Fig.~\ref{fig:monqwf}. Note that both modulation eigenvalues \eqref{eq:monatmod} have to be non-negative to be physical. This condition together with Eq.~(\ref{eq:monatin}) result in the threshold on the input energy
\begin{equation}
	\overline{n} \geq \overline{n}_\mathrm{thr} = \frac{1}{2}\left( \sqrt{\frac{\max{\{\gamma_\mathrm{env}^{(q)},\gamma_\mathrm{env}^{(p)}\}}}{\min{\{\gamma_\mathrm{env}^{(q)},\gamma_\mathrm{env}^{(p)}\}}}} + |\gamma_\mathrm{env}^{(q)} - \gamma_\mathrm{env}^{(p)}| - 1 \right),
\end{equation}
below which the obtained solutions do not hold.

Knowing all optimal eigenvalues of the system, we can easily determine the one-shot capacity defined by Eq.~(\ref{eq:ncapacity}),
\begin{equation}
  C_1 = g\left(\overline{n} + \frac{\gamma_\mathrm{env}^{(q)} + \gamma_\mathrm{env}^{(p)}}{2}\right) - g\left(\sqrt{\gamma_\mathrm{env}^{(q)} \, \gamma_\mathrm{env}^{(p)}}\right). \quad \overline{n} \geq \overline{n}_\mathrm{thr}.
  \label{eq:monatcap}
\end{equation}
One notices that in the first term the algebraic mean of the noise $(\gamma_\mathrm{env}^{(q)} + \gamma_\mathrm{env}^{(p)})/2$ and in the second term the geometric mean $[\gamma_\mathrm{env}^{(q)}\gamma_\mathrm{env}^{(p)}]^{-1/2}$ appears. Comparing $C_1$ with the definition of the Holevo quantity \eqref{eq:holevo} one can say that the first term corresponds to the entropy of a thermal state with a mean photon number identical to the mean energy of the system, and, roughly speaking, the second term corresponds to the average of the entropies of two thermal states with means $\gamma_\mathrm{env}^{(q)},\gamma_\mathrm{env}^{(p)}$. In the case $\gamma_\mathrm{env}^{(q)} = \gamma_\mathrm{env}^{(p)} = N$ we recover the capacity of a mono-modal thermal channel $C = g(\overline{n} + N) - g(N)$.

\section{\label{sec:mark}Gauss-Markov channel}

Now we proceed with the central part of this paper: the treatment of the multimode channel as introduced in Eq.~(\ref{eq:changen}) with a Gauss-Markov noise.

\subsection{The noise model}
Assume a Gaussian distributed real random vector $\bm{Z}$ that is generated by a Markov process:  
\begin{equation}
  	Z_i = \phi \, Z_{i-1} + W_i, \quad 0 \leq \phi < 1,
	\label{eq:gaussmark}
\end{equation}
where $W_i$ are Gaussian distributed and identically and independently distributed (i.d.d.). We set without loss of generality the expectation values $\mathrm{E}(Z_i) = \mathrm{E}(W_i) = 0$. We request in addition that the variance $\mathrm{Var}(Z_i) = N$, which leads to $\mathrm{Var}(W_i) = (1-\phi^2)N$. Then we conclude that the matrix elements of the covariance matrix $\bm{M}$ for the Gauss-Markov process \eqref{eq:gaussmark} read
\begin{equation}
  M_{ij}(\phi) \equiv \mathrm{Cov}{(Z_i,Z_j)} = N\phi^{|i-j|}, \quad 0 \leq \phi < 1,
  \label{eq:markmat}
\end{equation} 
where $N \in \mathbb{R}$ denotes the variance and $\phi$ is the nearest neighbor correlation. Equation \eqref{eq:markmat} is a symmetric Toeplitz matrix which is completely defined by its diagonal elements $t^{(M)}_k = N \phi^{|k|} = t^{(M)}_{-k}$ (see Appendix \ref{sec:appCirc}).

In the following, we treat the channel given by Eq.~(\ref{eq:changen}) with a classical noise covariance matrix
\begin{equation}
  \bm{\gamma_\mathrm{env}} = 
  \begin{pmatrix}
  \bm{M}(\phi) & 0\\
	0 & \bm{M}(-\phi)
	\end{pmatrix},
	\label{eq:marknoise}
\end{equation}
where $\bm{M}(\phi)$ reads as in Eq.~(\ref{eq:markmat}). In other words, the noise of the q (p) quadratures results from a Gauss-Markov process with (anti-)correlated nearest neighbors.

\subsection{\label{sec:solution}Solution}

In order to determine the capacity of the memory channel \eqref{eq:changen}, one needs to find the optimal input and modulation covariance matrices which obey the energy constraint \eqref{eq:enconstrcov}. One can treat equally this problem in the basis where the channel becomes memoryless provided that one can rotate the noise covariance matrix by a passive symplectic transformation into the basis where it is diagonal. This is possible, because a passive symplectic transformation does neither change the entropy, nor the energy constraint of the system. For the class of additive bosonic channels, where the noise covariance matrix $\bm{\gamma_\mathrm{env}}$ does not contain any cross correlations between the quadratures, i.e. has the form
\begin{equation}
	\bm{\gamma_\mathrm{env}} = 
	\begin{pmatrix} \bm{\gamma_\mathrm{env}}^{(q)} & 0\\0 & \bm{\gamma_\mathrm{env}}^{(p)} \end{pmatrix},
	\label{eq:noiseblock}
\end{equation}
one can show, that the transformation that diagonalizes the noise is symplectic only if the two block matrices $\bm{\gamma_\mathrm{env}}^{(q)}$ and $\bm{\gamma_\mathrm{env}}^{(p)}$, respectively, commute (see appendix \ref{sec:appCirc}).

Applied to the Gauss-Markov noise $\bm{\gamma_\mathrm{env}}$ as defined in Eq.~\eqref{eq:marknoise}, this condition requires that $[\bm{M}(\phi),\bm{M}(-\phi)] = 0$. However, for finite $n$ one can show that this does not hold and, as a consequence, $\bm{\gamma_\mathrm{env}}$ cannot be diagonalized by a passive symplectic transformation. In order to overcome this problem, we introduce another noise covariance matrix $\bm{\gamma_\mathrm{env}'}$ that does have commuting block matrices (see \eqref{eq:noiseblock}) and, furthermore, an eigenvalue spectrum that asymptotically converges to the eigenvalue spectrum of $\bm{\gamma_\mathrm{env}}$. The capacity of this modified channel system and of the original one will be identical, since this quantity is found in the limit of an infinite number of channel uses.

We introduce the noise covariance matrix   
\begin{equation}
  \bm{\gamma_\mathrm{env}}' = 
  \begin{pmatrix}
  \bm{M}^{(C)}(\phi) & 0\\
	0 & \bm{M}^{(C)}(-\phi)
	\end{pmatrix},
	\label{eq:circnoise}
\end{equation}
where $\bm{M}^{(C)}(\phi)$ is a circulant symmetric matrix 
\begin{equation}
	M^{(C)}_{ij}(\phi) = 
  	\left\{ 
		\begin{array}{l} 
      	\phi^{|i-j|}, \qquad \, 0 < |i-j| \leq \kappa \\
      	\phi^{n-|i-j|}, \quad \kappa < |i-j| \leq n-1, 
		\end{array} 
	\right. 
	\label{eq:circmat}
\end{equation}
with $\kappa = (n-1)/2$ for odd $n$ and $\kappa = n/2$ for even $n$, and $i,j = 1,...,n$. It is known that all circulant symmetric matrices commute, and furthermore, that the spectrum of a symmetric Toeplitz matrix and a circulant symmetric matrix generated by the same diagonals asymptotically coincides (see appendix \ref{sec:appCirc} for further details).

At this point, we base the following calculation on the conjecture that the optimal input covariance matrix $\bm{\gamma_\mathrm{in}}'$ diagonalizes in the same basis as the noise covariance matrix $\bm{\gamma_\mathrm{env}}'$. We rotate the matrix $\bm{\gamma_\mathrm{env}}'$ into the basis where it is diagonal and therefore treat a set of $n$ uncorrelated channels. Then, for each channel the optimal input takes the form derived in Eq.~(\ref{eq:monatin}), provided the total input energy suffices to exceed the threshold for each channel. We note in addition, that by Eq.~(\ref{eq:nchi}) and the subadditivity of the entropy
\begin{equation}
  S(\bm{\overline{\rho}}) \leq S(\overline{\rho}_1) + S(\overline{\rho}_2) + ... + S(\overline{\rho}_n),
  \label{eq:modsubadd}
\end{equation}
where 
\begin{equation}
	\overline{\rho}_k = \mathrm{Tr}_{\overline{\rho}_1,...,\overline{\rho}_{k-1},\overline{\rho}_{k+1},...\overline{\rho}_n}(\bm{\overline{\rho}}),
\end{equation}
the optimal modulation covariance matrix $\bm{\gamma_\mathrm{mod}}'$ is also diagonal in this basis.

We now take the limit $n \rightarrow \infty$. In appendix \ref{sec:appToep}, we determine the spectrum of $\bm{M}(\phi)$ in this limit, which is identical to the asymptotic spectrum of $\bm{M}^{(C)}(\phi)$. Thus, as the spectra of the two noise matrices $\bm{\gamma_\mathrm{env}}$ and $\bm{\gamma_\mathrm{env}}'$ are now identical, the optimal input and modulation covariance matrices $\bm{\gamma_\mathrm{in}}$ and $\bm{\gamma_\mathrm{in}}'$, ($\bm{\gamma_\mathrm{mod}}'$ and $ \bm{\gamma_\mathrm{mod}}'$) coincide.

In this limit the spectrum of $\bm{\gamma}_\mathrm{env}$ as introduced in Eq.~(\ref{eq:marknoise}) becomes continuous and reads for the two quadrature blocks
\begin{equation} 
	\gamma_\mathrm{env}^{(q)}(x) = \gamma_\mathrm{env}^{(p)}(\pi-x) = \lambda^{(M)}(x),
\end{equation}
where $\lambda^{(M)}(x)$ is the spectrum of $\bm{M}(\phi)$ given by Eq.~(\ref{eq:markspectrum}), i.e.
\begin{equation}
  \lambda^{(M)}(x) = N \, \frac{1 - \phi^2}{1 + \phi^2 - 2\phi \cos(x)},
  \label{eq:evn}
\end{equation}
where $x \in [0,\pi]$ is the spectral parameter.

For each individual channel, we know from Eq.~(\ref{eq:monthermal}) that if its allocated input energy is sufficient then the optimal overall output state is a thermal state, i.e.
\begin{equation}
  \gamma_\mathrm{in}^{(q)}(x) + \gamma_\mathrm{env}^{(q)}(x) + \gamma_\mathrm{mod}^{(q)}(x) = \gamma_\mathrm{in}^{(p)}(x) + \gamma_\mathrm{env}^{(p)}(x) + \gamma_\mathrm{mod}^{(p)}(x).
  \label{eq:markthstate}
\end{equation}
As in the mono-modal case each input has to match the anisotropy of the corresponding noise, i.e.
\begin{equation}
  \gamma_\mathrm{in}^{(q,p)}(x) = \frac{1}{2}\sqrt{\frac{\gamma_\mathrm{env}^{(q,p)}(x)}{\gamma_\mathrm{env}^{(p,q)}(x)}}.
  \label{eq:markin}
\end{equation}

In order to check whether the optimal input state is entangled in the original basis (i.e. the basis where the noise covariance matrix is introduced in \eqref{eq:marknoise}), we use  the transformation $\bm{Q}$ given by \eqref{eq:Q} for each quadrature block, to rotate the covariance matrix of the state back. In this basis, we conclude that the first mode is in a thermal state for correlations $\phi > 0$ (a coherent state for $\phi = 0$), with variances 
\begin{equation}
	\widetilde{\gamma}_\mathrm{in,1}^{(q)} = \widetilde{\gamma}_\mathrm{in,1}^{(p)} = \frac{1}{2\pi}\int\limits_0^{\pi}{d x \sqrt{\frac{1+\phi+2\phi\cos{(x)}}{1+\phi-2\phi \cos{(x)}}}}.
\end{equation}
Since the purity of the overall input state is untouched by the rotation, we conclude that the first mode is entangled with the rest of the modes. A full proof of entanglement could possibly be obtained by the same method applied to all modes.

We define the fraction of input energy used to prepare the squeezed (or entangled) input by
\begin{equation}
	\begin{split}
  \eta \equiv & \frac{1}{2\pi \overline{n}}\int\limits_0^{\pi}{d x \left(\gamma_\mathrm{in}^{(q)}(x) + \gamma_\mathrm{in}^{(p)}(x)\right)} - \frac{1}{2\overline{n}}\\ 
		    = & \frac{1}{\pi \overline{n}}\left(\int\limits_0^\pi{d x \; \gamma_\mathrm{in}^{(q)}(x)} - \frac{\pi}{2}\right),
	\end{split}
  \label{eq:marketa}
\end{equation}
where $\overline{n}$ is the mean maximum photon number constraint defined in Eq.~(\ref{eq:enconstr}). 

\begin{figure}
	\centering
		\includegraphics[width=0.45\textwidth]{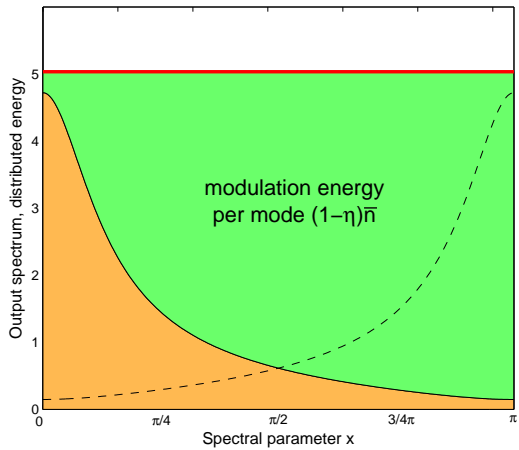} 
	\caption{(Color online) Global quantum water-filling solution: Output spectra $\gamma_\mathrm{out}^{(q)}(x)$ and $\gamma_\mathrm{out}^{(p)}(x)$ shown by the solid decreasing curve and dashed increasing curve vs. spectral parameter $x$. The solid bar represents the global quantum water-filling level $\mu_\mathrm{gl}$; the light gray area shows the energy used for modulation in $q$.}
	\label{fig:qwfmark}
\end{figure}

From classical information theory, we know that the amount of classical information sent through a system of parallel channels is maximized by a water-filling solution \cite{cove91}. By Eq.~\eqref{eq:markthstate} we confirm that this holds for the optimal modulation spectrum, i.e.
\begin{equation}
  \gamma_\mathrm{mod}^{(q,p)}(x) = \mu_\mathrm{gl} - \gamma_\mathrm{out}^{(q,p)}(x),
  \label{eq:markmod}
\end{equation}
where $\gamma_\mathrm{out}^{(q,p)}(x) = \gamma_\mathrm{in}^{(q,p)}(x) + \gamma_\mathrm{env}^{(q,p)}(x)$ is the (quantum) output spectrum and $\mu_\mathrm{gl}$ is now a global quantum water-filling level over all channels, such that
\begin{equation}
  \frac{1}{\pi}\int\limits_0^{\pi}{d x \, \gamma_\mathrm{mod}^{(q,p)}(x)} = (1-\eta)\overline{n}.
\end{equation}
The solution is graphically represented in Fig. \ref{fig:qwfmark}.
One concludes that the input energy threshold which determines whether the modulation \eqref{eq:markmod} is positive over the whole domain $x \in [0,\pi]$ is given by
\begin{equation} 
	(1-\eta)\overline{n} \geq \gamma_\mathrm{out}^{(q)}(0) - \frac{1}{\pi}\int\limits_0^\pi{d x \; \gamma_\mathrm{out}^{(q)}(x)}.
	\label{eq:markwcond1}
\end{equation}
Using the definition in Eq.~(\ref{eq:marketa}) and the property of the environment
\begin{equation}
	\frac{1}{\pi}\int\limits{d x \; \gamma_\mathrm{env}^{(q,p)}(x) = N},
	\label{eq:markintenv}
\end{equation}
we define
\begin{equation}
  \overline{n}_\mathrm{thr}(\phi,N) = \left(\frac{1+\phi}{1-\phi} - 1\right)\left(N + \frac{1}{2}\right),
  \label{eq:marknmthr}
\end{equation}
i.e. the minimum input energy needed to modulate the whole channel spectrum. 

If we assume that for given $\phi, N$
\begin{equation}
	\overline{n} \geq \overline{n}_\mathrm{thr}(\phi,N),
\end{equation}
then (see Fig \ref{fig:qwfmark}), using Eq.~(\ref{eq:marketa}) and Eq.~(\ref{eq:markintenv}), we determine the global quantum water-filling as
\begin{equation}
  \begin{split}
  \mu_\mathrm{gl} & = (1-\eta)\overline{n} + \frac{1}{\pi}\int\limits_0^\pi{dx \; \gamma_\mathrm{out}^{(q)}(x)}\\
      & = \overline{n} + N + \frac{1}{2},
  \end{split}
  \label{eq:markmu}
\end{equation}	
which corresponds to the overall modulated output variance of each of the channels, i.e.
\begin{equation}
  \overline{\gamma}^{(q)}(x) = \overline{\gamma}^{(p)}(x) = \mu_\mathrm{gl} = \overline{n} + N + \frac{1}{2},
  \label{eq:markcmix}
\end{equation}
where $\overline{\gamma}^{(q,p)}(x)$ correspond to the spectra of the covariance matrix $\bm{\overline{\gamma}}$ as defined in Eq.~(\ref{eq:chancov}). With the definition of the symplectic spectra for the quantum output, the total output and the noise in the asymptotic limit, that is
\begin{equation}
  \begin{split}
  	\nu_\mathrm{out}(x) & = \sqrt{\gamma_\mathrm{out}^{(q)}(x) \, \gamma_\mathrm{out}^{(p)}(x)}\\
  	\overline{\nu}(x) & = \sqrt{\overline{\gamma}^{(q)}(x) \,\overline{\gamma}^{(p)}(x)}\\
  	\nu_\mathrm{env}(x) & = \sqrt{\gamma_\mathrm{env}^{(q)}(x) \, \gamma_\mathrm{env}^{(p)}(x)},
  \end{split}
  \label{eq:marksympl}
\end{equation}
we are now ready to determine the capacity of the channel above threshold for an infinite number of uses: we take the limit of Eq.~(\ref{eq:ncapacity}) where the sum tends to an integral over the whole domain, i.e.
\begin{eqnarray}
  	C   & = & \lim_{n \rightarrow \infty} \frac{1}{n}\sup_{\bm{\gamma_\mathrm{in}},\bm{\gamma_\mathrm{mod}}} \, \chi_n \nonumber\\
    	& = & \frac{1}{\pi}\int\limits_0^\pi{dx \, \left\{ g\left(\overline{\nu}(x) - \frac{1}{2}\right) - g\left(\nu_\mathrm{out}(x) - \frac{1}{2}\right) \right\} } \nonumber\\
    	& = &  g(\overline{n} + N) - \frac{1}{\pi}\int\limits_0^\pi{dx \; g\left(\sqrt{\gamma_\mathrm{env}^{(q)}(x) \, \gamma_\mathrm{env}^{(p)}(x)}\right)}.\\
 		& = &  g(\overline{n} + N) - \frac{1}{\pi}\int\limits_0^\pi{dx \; g\left(\nu_\mathrm{env}(x)\right)}, \quad \label{eq:markcapC}
\end{eqnarray}
where $\nu_\mathrm{env}(x)$ is the asymptotic symplectic spectrum of $\bm{\gamma_\mathrm{env}}$. We conclude that the capacity is given by the difference of the entropy of a thermal state with mean photons $\overline{n} + N$ and, roughly speaking, the mean entropy of the environment. This is a generalization of the capacity found for the mono-modal channel found in Eq.~\eqref{eq:monatcap}.

\section{Limiting cases}
In this section we analyze the capacity \eqref{eq:markcapC} in several limiting cases, such as the case of the classical limit, the limit of full correlations, the case of a symmetric noise model and the transition from a finite to an infinite number of channel uses.	

\subsection{Classical limit}
In the classical limit we increase the input energy and the noise while keeping the signal to noise ratio constant. From the limiting behavior of $g(x)$:
\begin{equation}
	\lim_{x \rightarrow \infty}[g(x) - \log(x)] = 0, 
\end{equation} 
we conclude that we have to replace the $g(x)$-functions in Eq.~\eqref{eq:markcapC} by logarithms. The integral term can be simplified with the help of Ref.~\cite{grad80} 
\begin{equation}
  \frac{1}{\pi}\int\limits_0^\pi{dx \; \log{\left(\sqrt{\gamma_\mathrm{env}^{(q)}(x) \, \gamma_\mathrm{env}^{(q)}(x)}\right)}} = \log{(N(1-\phi^2))}.
\end{equation}
Thus, we conclude that
\begin{equation}
  \lim_{\overline{n},N \rightarrow \infty, \; \overline{n}/N = c}C \equiv C_\mathrm{cl} = \log{\left(\frac{1}{1 - \phi^2}\left(1 + \frac{\overline{n}}{N}\right)\right)}.
  \label{eq:classlim}
\end{equation}
Since $g(x) < \log{(x)}$ we reach this limit from below and hence, for finite $\overline{n},N$, the classical capacity of the bosonic channel is always smaller than the capacity of the classical Gaussian additive channel. This is the expected result because for the bosonic channel (for $\phi \neq 0$) a certain amount of energy is needed to prepare the input squeezed state, which is not the case for the classical channel.
In the classical limit, the relative fraction of energy used for squeezing vanishes to zero and therefore the capacities coincide. 
\begin{figure}
	\centering
		\includegraphics[width=0.5\textwidth]{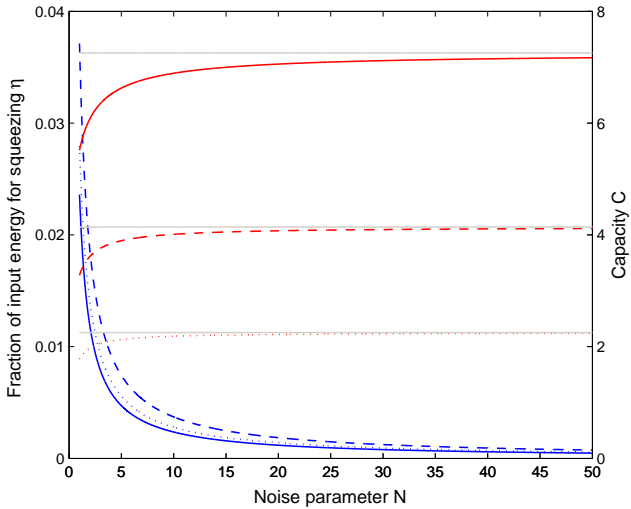}
	\caption{(Color online) Fraction of input energy used for squeezing $\eta$ and Capacity $C$ in bits per use vs. noise parameter $N$. The increasing curves correspond to the capacities, the decreasing to $\eta$. The gray horizontal lines correspond to the classical limit $C_\mathrm{cl}$ of each capacity. For all graphs the dotted, dashed and solid line correspond to $\phi = 0.4, 0.7, 0.9$. For each $\phi$ we set $\forall N \geq 1$: $\overline{n}/N = \overline{n}_\mathrm{thr}(\phi,N')/N'$ with $N'\equiv 1$ .}
	\label{fig:mark_classlim}
\end{figure}
In Fig. \ref{fig:mark_classlim} we plotted the capacity $C$ and relative fraction $\eta$ versus the noise parameter $N$. For each $\phi$ we fixed the signal to noise ratio $\overline{n}/N \equiv \overline{n}_\mathrm{thr}(\phi,N')/N',$ with $N' \equiv 1$ which guarantees to be above threshold $\forall N > 1$. The capacities all tend to their classical limit \eqref{eq:classlim}, while all fractions $\eta$ decrease towards zero. Interestingly, for fixed signal to noise ratio, we find that $\eta(\phi = 0.7) > \eta(\phi = 0.9)$. This is due to the fact that $\overline{n}_\mathrm{thr}(\phi=0.7,1) \ll \overline{n}_\mathrm{thr}(\phi=0.9,1)$. Although we need more input energy to match the higher correlation its cost with respect to the total energy $\overline{n}$ is lower.

\subsection{Full correlations}
We now investigate the limit of full correlations, i.e. $\phi \rightarrow 1$. Since the first term in Eq.~(\ref{eq:markcapC}) is independent of $\phi$ we take only the limit of the integral term (where we drop the constant $N$ inside the eigenvalue functions for now). For the integrand we find that
\begin{equation}
  \lim_{\phi \rightarrow 1}{g\left(\frac{1-\phi^2}{\sqrt{(1+\phi^2)^2 - 4 \, \phi^2 \, \cos^2{x}}}\right)} \rightarrow 0, \quad 0 < x < \pi.
  \label{eq:philim1}
\end{equation}
At the borders $x = \{0,\pi\}$ the integrand is equal to one for arbitrary $\phi$. As this contribution is however infinitesimally small it is easy to show that the integral in Eq.~(\ref{eq:markcapC}) vanishes and we conclude that
\begin{equation}
  \lim_{\phi \rightarrow 1}{C} = g(\overline{n} + N).
  \label{eq:markcapphilim}
\end{equation}
This result is in contrast to the result in the classical limit \eqref{eq:classlim}, for which
\begin{equation}
  \lim_{\phi \rightarrow 1}{C_\mathrm{cl}} \rightarrow \infty.
\end{equation}
Note that Eq.~\eqref{eq:markcapphilim} is valid only when the waterfilling condition \eqref{eq:marknmthr} is satisfied. However, as $\phi$ gets close to one while $\overline{n}$ is fixed, at some point this condition will be violated. In order to satisfy this condition with increasing $\phi$ one has to increase $\overline{n}$ according to \eqref{eq:marknmthr}, so that \eqref{eq:markcapphilim} becomes correct only as an asymptotic limit, with the right hand side diverging as $\phi$ goes to one. Nevertheless, we conjecture that for fixed $\overline{n}$ the capacity of the quantum channel will be finite in the limit $\phi \rightarrow 1$, because even in the absence of noise the capacity is finite as it is equal to $g(\overline{n})$. Hence, unlike in the classical case, for our quantum channel there is no diverging benefit from increasing correlations. This conjecture will be studied further in a forthcoming paper.

\subsection{Symmetric correlations}
In this subsection, we consider a modified, symmetric environment with same correlations in both quadratures:
\begin{equation}
  \bm{\gamma_\mathrm{env,s}} = 
  \begin{pmatrix}
  \bm{M}(\phi) & 0\\
	0 & \bm{M}(\phi)
	\end{pmatrix}.
	\label{eq:marknoises}
\end{equation}
As both quadrature blocks have now identical spectra, we treat a set of independent thermal channels, when $\bm{\gamma_\mathrm{env,s}}$ is diagonalized. We recover from Eq.~\eqref{eq:markin} immediately that in the asymptotic limit the optimal input spectra read
\begin{equation}
  \gamma_\mathrm{in,s}^{(q,p)}(x) = \frac{1}{2}\sqrt{\frac{\gamma_\mathrm{env,s}^{(q,p)}(x)}{\gamma_\mathrm{env,s}^{(p,q)}(x)}} = \frac{1}{2}, \quad \forall x,
  \label{eq:markins}
\end{equation}
i.e. the overall optimal input state is a set of coherent states and entanglement does not improve the transmission rate. This is in agreement with previous investigations of the two-mode model discussed in \cite{cerf05}. Recently, the same observation was made independently in Ref.~\cite{lupo09b}. 

\subsection{From finite to infinite number of uses}	
If one applies the method introduced in section \ref{sec:solution} to a noise covariance matrix, where the two quadrature block matrices (see \eqref{eq:noiseblock}) commute for finite $n$, then the optimal transmission rate reads as in \eqref{eq:markcapC}, where the integral is replaced by a sum term normalized by $n$:
\begin{equation}
	R^{(n)} = g(\overline{n} + N) - \frac{1}{n}\sum\limits_{k=1}^n{g\left(\sqrt{\gamma^{(q)}_{\mathrm{env},k} \, \gamma^{(p)}_{\mathrm{env},k}}\right)},
	\label{eq:R}
\end{equation}
where $\gamma^{(q)}_{\mathrm{env},k},\gamma^{(p)}_{\mathrm{env},k}$ denote the eigenvalue spectra for the given environment.

\begin{figure}
		\includegraphics[width=0.5\textwidth]{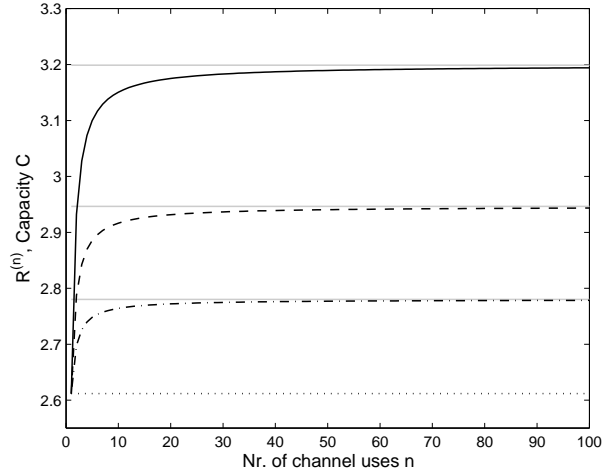}
	\caption{Function $R^{(n)}$ and capacity $C$ in bits per use vs. number of channel uses $n$. The dotted, dashed, dashed-dotted and solid curve correspond to $R^{(n)}$ with $\phi = 0,0.4,0.55,0.7$. For each $\phi$ the corresponding capacity is shown by a gray horizontal line. For all plots we took $N = 1$ and $\overline{n} = 7.5 > \overline{n}_\mathrm{thr}(\phi=0.7,N=1)$.}
	\label{fig:mark_nCapvsn}
\end{figure}

Although, for finite $n$ the two block matrices $\bm{M}(\phi)$ and $\bm{M}(-\phi)$ of the introduced Gauss-Markov noise \eqref{eq:marknoise} do not commute, we investigate $R^{(n)}$ as it converges with increasing channel uses $n$ to the capacity \eqref{eq:markcapC}, where $\gamma^{(q)}_{\mathrm{env},k},\gamma^{(p)}_{\mathrm{env},k}$ are the spectra of $\bm{M}(\phi),\bm{M}(-\phi)$ and numerically obtained. In Fig. \ref{fig:mark_nCapvsn} we plot $R^{(n)}$ for fixed $N$, various correlation strength $\phi$ and a fixed input energy $\overline{n}$, which is above threshold for the strongest $\phi$. In addition we denoted by gray bars for each $\phi$ the asymptotic capacity given by Eq.~(\ref{eq:markcapC}). In the plotted region, we observe that $R^{(n)}$ indeed converges to the capacity $C$. 
	
\section{Conclusions}
The classical capacity of a multi-mode channel with Gauss-Markov noise was found under certain assumptions above a certain input energy threshold. By diagonalizing the noise covariance matrix one was led to treat first the capacity of a one mode phase-dependent channel.

In the one-mode case we have shown that above the threshold the optimal quantum input state is a squeezed state where the squeezing matches the anisotropy of the noise. Furthermore, we found that the optimal overall modulated output state is a thermal state and therefore the classical modulation is determined by a ``quantum water-filling'' level. 

For the multi-mode channel with Gauss-Markov noise the optimal input and modulation were discussed in the asymptotic limit. Above an energy threshold, the optimal input eigenvalue of each channel is similar to the mono-modal solution determined by the anisotropy of the noise of the channel. The optimal modulation was found by a global quantum water-filling, which led us to the conclusion that the overall modulated output state is the same thermal state for all channels or equivalently all channel uses. When we rotated the covariance matrix of the overall input state back to the original basis, we confirmed that the first mode is in a thermal state and hence the total state is entangled.

Finally, several limit cases were discussed. First, the classical limit of the channel was discussed and we showed that the quantum expression tends to the classical expression. Secondly, we argued that in the limit of full correlations the capacity of the quantum model stays finite whereas it is diverging for its classical counterpart. In the case of a symmetric noise model, we recovered that the optimal input is given by a set of coherent states. At last we depicted the asymptotic behavior of a function which represents the optimal transmission rate for finite uses for commuting quadrature block matrices, and observe that it tends to the capacity for infinite channel uses. 

In addition we remark that the proposed solution method can be applied more generally to an environment which is constructed by a stationary Gauss processes, because these processes are characterized by symmetric Toeplitz matrices which commute asymptotically (see the appendix \ref{sec:appCirc} for more details). This means that for any pair of Toeplitz matrices $\bm{T,T'}$ replacing $\bm{M}(\phi),\bm{M}(-\phi)$ in \eqref{eq:marknoise} the capacity (above threshold) reads as in \eqref{eq:markcapC} where the noise spectra are replaced by the spectra of $\bm{T,T'}$.

\begin{acknowledgments}
NJC thanks Jeffrey H. Shapiro for useful discussions. JS thanks Ra\'{u}l García-Patr\'{o}n for helpful comments to the one-mode channel solution. We thank Oleg V. Pilyavets for  stimulating comments. We acknowledge financial support from the EU under projects COMPAS, from the Belgian federal program PAI under project Photonics@be and from the Brussels Capital region under the projects CRYPTASC and Prospective Research for Brussels program.
\end{acknowledgments}

\appendix

\section{\label{sec:appCirc}Toeplitz and circulant symmetric matrices}
In this section we specify the symplectic transformation that diagonalizes the noise $\bm{\gamma_\mathrm{env}}'$ \eqref{eq:circnoise} for finite dimension $n$ and $\bm{\gamma_\mathrm{env}}$ \eqref{eq:marknoise} for $n$ tending to infinity.
 
A quadratic matrix $\bm{T}$ with dimension $n \times n$ is called Toeplitz \cite{gray06} if
\begin{equation}
  T_{ij} = t^{(T)}_{i-j},
  \label{eq:toep}
\end{equation} 
where $t^{(T)}_{i-j}$ is a series of real numbers and $i,j = 1,...,n$. The matrix in Eq.~(\ref{eq:toep}) belongs to the Wiener class, if $\{t^{(T)}_k\}$ is absolutely convergent, that is,\\ $\sum_{k=-\infty}^\infty{|t^{(T)}_k|} < \infty$. Then the Fourier series 
\begin{equation}
	f_T(x) = \sum_{k=-\infty}^\infty{t^{(T)}_k e^{ikx}}, \quad x \in [0,2\pi]
	\label{eq:fourier}
\end{equation}
exists and is Riemann integrable. 

A Toeplitz matrix $\bm{T}^{(C)}$ of dimension $n$ is called circulant symmetric \cite{full96} if it has the form
\begin{equation}
  \bm{T}^{(C)} =
	\begin{pmatrix}
		t_0 		& t_1 		& t_2		& \cdots	& t_2		& t_1\\[5pt]
		t_1			& t_0		& t_1		& \cdots	& t_3		& t_2\\[5pt]
		t_2			& t_1		& t_0		& \cdots	& t_4		& t_3\\[5pt]
		\vdots		& \vdots	& \vdots	& 			& \vdots	& \vdots\\[5pt]
		t_1			& t_2		& t_3		& \cdots	& t_1		& t_0
	\end{pmatrix}.
  \label{eq:circ}
\end{equation}
The unitary transformation $\bm{Q}$, that diagonalizes $\bm{T}^{(C)}$ with odd dimension $n$, reads
\begin{equation}
	\begin{split}
	& \bm{Q}^{\ensuremath{\mathsf{T}}} = \sqrt{\frac{2}{n}} \times \\
	& \begin{pmatrix}
		\frac{1}{\sqrt{2}}		&	\frac{1}{\sqrt{2}}					&	\frac{1}{\sqrt{2}}					&	\cdots		&	\frac{1}{\sqrt{2}}\\[5pt]
		1						&	\cos{(\frac{2\pi}{n})}				&	\cos{(\frac{4\pi}{n})}				&	\cdots		&	\cos{(\frac{2\pi(n-1)}{n})}\\[5pt]
		0						& 	\sin{(\frac{2\pi}{n})}				&	\sin{(\frac{4\pi}{n})}				&	\cdots		&	\sin{(\frac{2\pi(n-1)}{n})}\\[5pt]
		1						&	\cos{(\frac{4\pi}{n})}				&	\cos{(\frac{8\pi}{n})}				&	\cdots		&	\cos{(\frac{4\pi(n-1)}{n})}\\[5pt]
		\vdots					&	\vdots								&	\vdots								&				& 	\vdots\\[5pt]
		0						& 	\sin{(\frac{n-1}{2}\frac{2\pi}{n})}	&	\sin{(\frac{n-1}{2}\frac{4\pi}{n})}	&	\cdots		&	\sin{(\frac{n-1}{2}\frac{2\pi(n-1)}{n})}
	\end{pmatrix}.
	\end{split}
	\label{eq:Q}
\end{equation}
For even $n$ the same pattern holds, except for a row $n^{-1/2}(1,-1,1,...,-1)$ at $i=n/2$. The diagonal matrix $\bm{Q}^{\ensuremath{\mathsf{T}}} \, \bm{T}^{(C)} \, \bm{Q}$ converges with increasing $n$ to $\bm{D}$, where 
\begin{equation}
	\bm{D} = \mathrm{diag}(d_1,d_2,...,d_n),
\end{equation}
where for odd $n$ 
\begin{equation}
	\begin{split}
		d_1 & = f_T(0)\\
		d_{2j} = d_{2j+1} & = f_T(2\pi k j/n),
	\end{split}
\end{equation}
$d_n = f_T(\pi)$ for even $n$, where $j = 1,2,...,(n-1)/2$ and $f_T(x)$ as defined in \eqref{eq:fourier}. In addition, if a Toeplitz matrix $\bm{T}$ belongs to the Wiener class, then $\bm{Q}^{\ensuremath{\mathsf{T}}} \, \bm{T} \, \bm{Q}$ also converges to $\bm{D}$. This means that $\bm{T}^{(C)}$ and $\bm{T}$ can be asymptotically diagonalized in the same basis, and hence that all Toeplitz matrices (which belong to the Wiener class) asymptotically commute.

Now we show, that the transformation, that diagonalizes $\bm{\gamma_\mathrm{env}}'$ as defined in \eqref{eq:circnoise} is indeed symplectic. A transformation $\bm{S}$ is called symplectic if
\begin{equation}
	\bm{S} \, \bm{J} \, \bm{S}^{\ensuremath{\mathsf{T}}} = \bm{J},
\end{equation}
where $\bm{J}$ is the commutation matrix defined in \eqref{eq:J}. For a covariance matrix $\bm{\gamma}$ of the shape
\begin{equation}
	\bm{\gamma} = \begin{pmatrix} \bm{\gamma}_q & 0\\0 & \bm{\gamma}_p \end{pmatrix},
\end{equation}
a rotation $\bm{R}$ that diagonalizes $\bm{\gamma}$ is of the shape 
\begin{equation}
	\bm{R} = \begin{pmatrix} \bm{U}_q & 0\\0 & \bm{U}_p \end{pmatrix}.
\end{equation}
Therefore, in order for $\bm{R}$ to be a symplectic transformation we find the requirement that
\begin{equation}
	\bm{U}_q^{\ensuremath{\mathsf{T}}} \, \bm{U}_p = \bm{U}_q \, \bm{U}_p^{\ensuremath{\mathsf{T}}} = \mathbb{1},
\end{equation}
where $\bm{U}_q,\bm{U}_p$ are orthogonal transformations. For the noise \eqref{eq:circnoise}, both quadrature blocks $\bm{\gamma}_q, \bm{\gamma}_p$ are diagonalized by the same orthogonal transformation $\bm{U}_q = \bm{U}_p = \bm{Q}$ and thus $\bm{R}$ corresponds to a passive symplectic transformation.

\section{\label{sec:appToep}Spectrum of a Gauss-Markov covariance matrix}

Our goal in the following is to determine the spectrum of $\bm{M}(\phi)$ defined in Eq.~(\ref{eq:markmat}). As this matrix generates a Markov process, one intuitively expects $\bm{M}^{-1}(\phi)$ to be three diagonal, such that only nearest neighbor terms appear in the exponential of the resulting Gaussian distribution.
Therefore we consider a matrix
\begin{equation}
  \bm{V} = \frac{1}{N}\,\frac{1 + \phi^2}{1 - \phi^2} \,
  \begin{pmatrix} 
  1                         & -\frac{\phi}{1 + \phi^2} & 0                        & 0      & \cdots \\ 
   -\frac{\phi}{1 + \phi^2} & 1                        & -\frac{\phi}{1 + \phi^2} & 0      & \cdots \\
  0                         & -\frac{\phi}{1 + \phi^2} & 1 & -\frac{\phi}{1 + \phi^2}      & \cdots \\
  0                         & 0                        & -\frac{\phi}{1 + \phi^2} & 1      & \ddots \\
  \vdots                    & \vdots                   &                          & \ddots & \ddots \\
  \end{pmatrix},     
  \label{eq:markinv}
\end{equation}
where $0 \neq N \in \mathbb{R}, 0 \leq |\phi| < 1$. We observe that
\begin{equation}
  \sum_{k=-\infty}^\infty{|t^{(V)}_k|} = \frac{1}{N} \frac{1+\phi^2}{1-\phi^2}\left(1+2\frac{|\phi|}{\phi^2+1}\right) < \infty,
\end{equation}
and therefore obtain the Fourier series 
\begin{equation}
  f_{V}(x) 
    = \frac{1}{N} 
      \frac{1+\phi^2}{1-\phi^2}
      \left(1+2\frac{|\phi|}{\phi^2+1} \cos(x) \right), 
  \label{eq:fourierV}
\end{equation}
where the right hand side of the latter was obtained by using \eqref{eq:fourier}. From \cite{gray06} we state that if the spectrum of a Toeplitz matrix $\bm{T}$ is strictly positive, then the inverse $\bm{T^{-1}}$ is asymptotically Toeplitz, with diagonals
\begin{equation}
  t_k^{(T^{-1})} = \frac{1}{2\pi}\int\limits_{-\pi}^\pi{dx \, \frac{e^{-ikx}}{f_T(x)}}, \quad n \rightarrow \infty.
  \label{eq:toepinv}
\end{equation}
Inserting \eqref{eq:fourierV} in the latter leads to
\begin{equation}
  t_k^{(V^{-1})} = N\phi^{|k|},
\end{equation}
or equivalently, we found that
\begin{equation}
	\lim_{n \rightarrow \infty}{\bm{V}} \rightarrow \bm{M}^{-1}(\phi).
\end{equation}
Hence, in the limit of infinite number of rows and columns we found the inverse of $\bm{M}(\phi)$. We determine the spectrum of $\bm{V}$ by using Ref.~\cite{grud05}, take its inverse and therefore receive the spectrum of $\bm{M}(\phi)$ in the limit of inifinte channel uses:
\begin{equation}
  \lambda^{(M)}(x)= N \, \frac{1 - \phi^2}{1 + \phi^2 - 2\phi \cos(x)}, 
  \label{eq:markspectrum}
\end{equation}
with the spectral parameter $x \in [0,\pi]$.

\newpage
\bibliography{publications-v10}

\end{document}